# Pull-off strength of mushroom-shaped fibrils adhered to rigid substrates

C. Betegón[a*], C. Rodríguez[a], E. Martínez-Pañeda[b], R.M. McMeeking[c,d,e]

[a]*Construction & Manufacturing Engineering Dept., University of Oviedo, Gijón 33203, Spain*
[b]*Dept. of Engineering Science, University of Oxford, UK*
[c]*Materials Dept. & Mechanical Engineering Dept., University of California, Santa Barbara CA 93106, USA*
[d]*Engineering School, University of Aberdeen, King's College, Aberdeen AB24 3UE, UK;*
[e]*INM – Leibniz Institute for New Materials, Campus D2 2, 66123 Saarbrücken, Germany.*

- Covadonga Betegón. Email: cova@uniovi.es

## Abstract

The exceptional adhesion properties of biological fibrillar structures -such as those found in geckos- have inspired the development of synthetic adhesive surfaces. Among these, mushroom-shaped fibrils have demonstrated superior pull-off strength compared to other geometries. In this study, we employ a computational approach based on a Dugdale cohesive zone model to analyze the detachment behavior of these fibrils when adhered to a rigid substrate. The results provide complete pull-off curves, revealing that the separation process is inherently unstable under load control, regardless of whether detachment initiates at the fibril edge or center.

Our findings show that fibrils with a wide, thin mushroom cap effectively reduce stress concentrations and promote central detachment, leading to enhanced adhesion. However, detachment from the center is not observed in all geometries, whereas edge detachment can occur under certain conditions in all cases. Additionally, we investigate the impact of adhesion defects at the fibril center, showing that they can significantly reduce pull-off strength, particularly at high values of the dimensionless parameter $\chi$. These insights contribute to the optimization of bio-inspired adhesives and microstructured surfaces for various engineering applications.

## Introduction

The shape of the fibrillar structures found on the feet of some animals, such as the gecko, has inspired the development of synthetic adhesive patterned surfaces [1-4]. The most commonly fabricated surface patterns are arrays of compliant pillars in the micro- and nanometer range. Over the last decades, researchers have proposed several hypotheses to explain the origin of the high adhesion strengths of these fibrillar structures. The associated studies tend to show that the effectiveness of their adhesion is mainly due to van der Waals forces [5,6], and that these may be enhanced by the presence of water vapor in the environment leading to an additional capillary effect[7]. Various mechanical models have been developed to represent specific hairy attachment systems within the framework of adhesive contact mechanics[8-11]. Autumn et al.[5] and Arzt et al.[12] identified one of the basic principles of adhesion of a fibrillar interface, that the strength increases as the size of each fibril is reduced because the stress to detach an individual fibril varies inversely with its size. This is referred to as contact splitting.

A considerable number of studies has been dedicated to finding an optimal shape and size for fibrils for patterned adhesive surfaces[11,13-18]. One fibril shape that has received extensive attention is the so-called "mushroom" geometry, in which the adhesive end widens into a circular flange. This geometry has been shown to improve adhesion compared to others, and a number of studies have been devoted to explaining its superior strength.[17, 19-21] By assuming a pre-existing defect at the edge of the adhesion area and Griffith's criterion applied to separation controlled by adhesion energy, Spuskanyuk et al.[17] analyzed detachment from a rigid surface of both a straight cylindrical punch fibril and a mushroom tipped one. They observed that the defect is much more damaging to the pull-off strength for the cylindrical punch and thus concluded that mushroom shaped fibrils are more adhesive. Carbone et al.[20] studied the detachment mechanism of both configurations, considering three different possible scenarios: (i) detachment propagation from the pillar edge; (ii) propagation of interfacial detachment defects from the interior; and (iii) detachment due to achievement of the theoretical van der Waals strength without the presence of pre-existing detachment defects. Their results revealed that, as determined by Spuskanyuk et al.[17], mushroom fibrils are superior to cylindrical ones in their ability to eliminate strong stress singularities, thereby diminishing the impact of edge defects at the interface. Carbone et al.[20] determined pull-off strengths by theoretical and computational means, and proposed maps for the detachment behavior of different fibril geometries.

Cohesive zone models have proven to be a suitable tool to determine the pull-off strength of both straight-sided punch and mushroom fibrils [4, 21-26], with the Dugdale form[27] being the most widely used. This model assumes that the detachment process initiates when the normal interfacial stress reaches the theoretical adhesion strength of the interface, denoted by $\sigma_0$. Separation continues at this stress until a critical gap width, $\delta_C$, is reached, after which the interface can no longer support traction, resulting in the initiation of complete detachment at the point in question. Complete detachment then spreads from the initiation location. Both $\sigma_0$ and $\delta_C$ are interface properties, and the work of adhesion of the surfaces in contact is given by $W_{adh} = \sigma_0 \delta_C$. The region where the separation of the interface occurs under tension is referred to as the cohesive or Dugdale zone.

Using this model, Tang et al.[22] showed that the average interface stress, $\bar{\sigma}$, needed to detach a compliant cylindrical fibril attached to a rigid substrate is governed by a single dimensionless parameter $\chi$ defined as

$$\chi = \frac{\sigma_0^2 D (1 - \nu^2)}{2\pi E W_{adh}} \qquad (1)$$

where $D$ is the fibril diameter, $E$ the elastic modulus of the fibril and $\nu$ its Poisson´s ratio. The parameter $\chi$ is the ratio of two lengths, the fibril radius $D/2$ and the length $b = \pi E W_{adh}/\sigma_0^2 (1 - \nu^2)$, which is an estimate of the size of the Dugdale cohesive zone upon detachment. The pull-off stress, equal to $\sigma_C$, is maximized and equal to the theoretical strength, $\sigma_0$, when $\chi \ll 1$, i.e., when the fibril is small in diameter compared to the length of the Dugdale cohesive zone at detachment that would prevail in a larger fibril. This situation is known as the flaw insensitive regime, as detachment defects, edge singularities and other features cannot reduce the detachment strength significantly, and this regime can be reached both for the cylindrical punch and the mushroom fibril if $\chi$ is small enough.

Aksak et al.[21] studied the pull-off strength of mushroom fibrils, identifying its dependence on $\chi$ and on the fibrillar geometry. In this case, the fibril shape consisted of a straight shaft with a tip in the shape of a truncated cone widening towards the adhesive end. Various conical angles and diameters of the shaft and the adhesive end were considered. Aksak et al.[21] also found that the location of detachment initiation, either at the edge or at the center of the fibril, depended on $\chi$ for a given geometry. Detachment at the edge implicates the effect of a stress concentration: the intensity of the concentration increases with the load up to a critical value that causes detachment to initiate and propagate until separation occurs. This mode of rupture occurs at a relatively small pull-off load since, in the circumstances involved, only a small fraction of the contact is subjected to high stresses. In contrast, when detachment starts at the center of the fibril, it occurs under conditions where the stress is more uniform in the contact zone, and, consequently, the pull-off load is greater. Therefore, the design of a fibrillar geometry able to invoke high pull-off loads must ensure that detachment starts at the center of the fibril. This requires a sound understanding of the factors that determine how detachment initiates for a given fibrillar geometry.

The competition between the two possible ways in which detachment can initiate for a mushroom fibril has been analyzed by Balijepalli et al.[28] By assuming a pre-existing edge defect having a given universal size, they predicted the adhesion for various mushroom cap shapes, revealing higher strengths for a smaller stalk diameter and a thin mushroom cap having a large diameter. They also concluded that detachment from the center of the fibril can be induced if the mushroom cap is effective enough. This phenomenon is favored by the introduction of a fillet radius, connecting the stalk to the cap of the mushroom fibril, that avoids stress concentrations being generated in the adhesive contact at the location just under the edge of the stalk. The presence of such a stress concentration induces pull-off at a load well below that prevailing when the stress on the adhesive contact is more uniform, leading to detachment initiation at the fibril center.

Both Aksak et al.[21] and Tang et al.[22] assume in their works that the adhesion between the fibril and the substrate is initially complete over the contact area. Nevertheless, there is strong experimental evidence[15, 29, 30] that this theoretical pull-off strength is reduced by the presence of adhesion defects at the interface and, as the defects act as stress raisers, their propagation will determine the fibril pull-off strength. These defects may result from surface roughness, fabrication imperfections or contaminant particles, it having been shown that the probabilistic variation in their size and location defines the statistical properties of fibril detachment strength[31].

In this work, the detachment of the mushroom shaped fibril suggested by Balijepalli et al.[28] is analyzed computationally, with a Dugdale cohesive zone model used to represent adhesion to a rigid substrate. As shown in figure 1(a), the fibril has stalk diameter $D$, mushroom flange diameter $D_f$, flange thickness $h$, fillet radius $R$ and stalk length $L$. Traction free conditions at the interface between the fibril and the substrate are generated only when the separation at the adhesion surface reaches the critical value $\delta_C$.

We rely on the analysis of Tang et al.[22] as the basis of our assumption that the pull-off strength depends on the parameter $\chi$ plus the geometric ratios $\beta = D_f/D$, $\xi = h/D_f$ and $R/D_f$. The Tang et al. [22] deductions, as given in their Appendix B, are reliable in this regard as long as pull-off coincides with the stage in the deformation process where the maximum separation in the Dugdale cohesive zone first attains the critical separation $\delta_C$. We therefore study this feature of the behavior to determine whether the Tang et al.[22] deduction can be exploited, and having done so, deduce that the pull-off strength of a fibril has the functional form

$$\frac{\sigma_C}{\sigma_0} = f\left(\chi, \frac{h}{D_f}, \frac{D_f}{D}, \frac{R}{D_f}\right) \tag{2}$$

where $f$ is a function of the parameters listed in the parenthesis, and where $\chi$ is now defined to be

$$\chi = \frac{\sigma_0^2 D_f (1-\nu^2)}{2\pi E W_{adh}} = \frac{\sigma_0 D_f (1-\nu^2)}{2\pi E \delta_C} \tag{3}$$

The present study has been structured as follows. First, the load-extension curves in the unstable detachment process are obtained for two mushroom fibril geometries, selected in such a way that in the first one the detachment starts from the center and in the second one it starts from the edge. Secondly, a systematic study is conducted of the influence of the mushroom cap dimensions on both the pull-off resistance and the detachment initiation site. The dependence of the pull-off behavior on the adhesion stress, $\sigma_0$, and the critical separation, $\delta_C$, is investigated. In these two studies, we assume that contact and adhesion between the fibril and the substrate are initially complete over the contact area, without any detachment defect or heterogeneity, and with no segment in the interface where tensile tractions are initially forbidden. In the third part of the study, we assume the existence of an adhesion defect at the center of the fibril within which cohesive stress across the interface is absent. We investigate the influence of this adhesion defect on the pull-off strength of those geometries in which the stress distribution favors the propagation of defects at the center of the fibril.

**Detachment process**

To understand the adhesion capacity of different mushroom shaped fibrils, the pull-off and associated detachment processes have been modeled using the finite element method. Fibrils with mushroom caps, as shown in figure 1(a), are subjected to boundary conditions and deformations that simulate detachment governed by Dugdale cohesive interaction; we consider an initial stress free state in the fibril before detachment and we also assume that the fibril surface does not displace latterly on the rigid surface. Thus, throughout the calculations, the radial displacements of all points on the bottom surface of the fibril are held equal to zero. The processes of attachment and detachment are modelled by first suspending a fibril from a stationary support at its top and applying a uniform tensile traction, $\sigma_0$, on the entire bottom surface of a fibril while at the same time lateral displacement of the bottom surface of the fibril is forbidden. Initially there is no contact between the fibril and the rigid substrate below it. This condition, but with tractions applied only to a segment of the bottom surface, is illustrated in figure 1(b). Note that in figure 1(b) the resulting deformation of the fibril is greatly exaggerated for clarity. In addition, we emphasize that the computations being described currently involve Dugdale tractions everywhere on the bottom surface of the fibril, so that $c = 0$. The case where $c \neq 0$ is described below.

Next, with the top of the fibril held fixed, the rigid substrate is raised and contact is made between it and the deformed lower surface of the fibril. Thereafter the rigid substrate continues to rise upwards. Contact conditions are enforced using the direct contact constraint method, causing the rigid substrate to deform the fibril from its shape of figure 1(b), resulting in the shapes shown schematically in figures 1(c) and 1(d). With the substrate continuing to move upward, deformation of the fibril continues until a stress free state in the axial direction is reached, with this final state in the finite element simulations corresponding to the initial state of the fibril prior to detachment. Thus, our computations simulate detachment of the fibril by a reversal of the process.

At each stage of the simulation the distance $L + \Delta$ between the top of the fibril and the substrate is recorded; this is shown in figures 1(c) and 1(d), with the parameter $\Delta$ representing the current extension of the fibril plus any gap between its bottom surface and the substrate. Similarly, the finite element results provide the upward load, $P$, applied to the top of the fibril. An important parameter that is also monitored is the distance $\delta_C$ between the rigid substrate and the lower surface of the fibril at $r = 0$. Recall that the case being described has $c = 0$. In some cases $\delta$ at the perimeter of the flange is greater than its value at $r = 0$, and in that case $\delta_C$ is defined to be the value of $\delta$ at the perimeter, i.e., at $r = D_f/2$.

The applied load, $P$, at any stage is computed from the results of the finite element computations, as noted above. Thus, for a given value of $\sigma_0$, the applied load is calculated as a function of $\Delta$, thereby providing the load deflection curve for the fibril under tension. However, the applied load can also be considered to be a function of $\delta_C$, and therefore of $\chi$. Since $\delta_C$ can be defined to be the critical separation beyond which Dugdale cohesion is not effective, $P$, taken to be a function of $\chi$, can be interpreted, when $c = 0$, as the pull-off load for that value of $\chi$, i.e., for the current value of $\delta_C$. We note one qualification, however, which is that this interpretation of $P$ as the pull-off load is only robust if pull-off is unstable and commences as soon as $\delta$ reaches $\delta_C$ on the bottom surface of the fibril during the process of separation of the fibril from the substrate.

To investigate whether pull-off becomes unstable when initiated at the center of the fibril, an additional set of calculations is carried out as follows. The same procedure as described above, is utilized, but with the uniform cohesive traction, $\sigma_0$, applied only on a portion of the bottom surface, i.e., now with $c \neq 0$. As before, no slip conditions are imposed everywhere between the fibril surface and the substrate, including in the traction free region. For geometries in which detachment initiates at the fibril center, the cohesive traction is applied only on the bottom surface where $r \geq c$, with $c$ being a fixed, non-zero radius. For these calculations, the parameter $\delta_C$ is now computed at $r = c$, as illustrated in figure 1(c). Since the Dugdale cohesive traction, $\sigma_0$, is applied only for $r \geq c$, the distance $\delta_C$ is, for cases where detachment commences at the center of the fibril, the critical distance beyond which the lower surface of the fibril does not interact cohesively with the rigid substrate. It follows that the deformation depicted in figure 1(c), and the values of the applied load, $P$, and the fibril displacement, $\Delta$, constitute a snapshot solution for an adhered fibril subject to Dugdale cohesion with traction $\sigma_0$, a central detached region of radius $c$, and a critical interaction distance $\delta_C$ at the edge of the detached region.

Since detachment can initiate at the fibril edge, calculations are also carried out, with no-slip conditions, for the setup illustrated in figure 1(d). In this case, the cohesive stress applied only on the bottom surface of the fibril where $r \leq D_f/2 - c$, with $c$ having a fixed value, and $\delta_C$ computed at $r = D_f/2 - c$. The deformation depicted in figure 1(d), and the values of the applied load, $P$, and the fibril displacement, $\Delta$, constitute a snapshot solution for an adhered fibril subject to Dugdale cohesion with traction $\sigma_0$, an external detached annulus of width $c$, and a critical interaction distance $\delta_C$ at the verge of the detached region.

For each fibrillar geometry, results are obtained for 200 different values of $c$, ranging from 0 to $D_f/2$, in a total of 4000 equilibrium solutions. To define the fibrillar geometry the cap diameter, $D_f$, is held fixed at 12.5 μm, with the undeformed fibril length chosen throughout to be $L = 30$ μm. Both the stalk diameter, $D$, and the cap thickness, $h$, are varied and the influence of $\beta = D_f/D$ and $\xi = h/D_f$ ascertained. Values of $\beta$= 1.05, 1.15, 1.2 and 1.25 are chosen, and for each of the values of $\beta$ three cap thicknesses are considered ($\xi$=0.01, 0.025 and 0.05). These values of $\beta$ and $\xi$ are chosen to represent realistic and practical geometries of mushrooms fibrils. Whenever possible a fillet radius $R = 0.083D_f$ is used between the stalk and the mushroom flange, as in Balijepalli et al. [28]. If such a fillet radius is too large for a complete 90° arc to be possible, $R$ (see figure 1(a)) is taken to be as large as possible consistent with the fillet having such a 90° arc.

The commercial finite element code Abaqus-Standard [32] is used with meshes ranging from 1,168,702 to 1,172,898 linear quadrilateral hybrid axisymmetric elements, with 216,000 square elements on the interface surface in all the meshes. Details regarding the mesh topology and the element size at the interface are provided in Appendix A. The chosen element size was selected to resolve the stress and separation gradients across the cohesive zone accurately. Although full global mesh refinement was computationally impractical due to the large number of elements, a localized refinement study was conducted for a representative case. The results confirmed that the mesh resolution is sufficient, as no significant differences were observed in the traction or separation fields near pull-off.

The calculations are carried out under the assumptions of linear, isotropic elasticity subject to conditions of infinitesimal strain. The resulting deformation is modeled as nearly incompressible by setting Poisson's ratio equal to 0.499, and Young's modulus equals 2 MPa. The value of the interface strength, $\sigma_0$, is taken as 0.1 MPa. For $\beta = 1.05$ and $\xi = 0.025$ the calculations are repeated with $\sigma_0 = 0.05$ MPa and $\sigma_0 = 0.2$ MPa to confirm that the results for the pull-off strength depend on $\chi, \beta, \xi$ and $R/D_f$ only, and that there is no dependence on the individual parameters within $\chi$.

We note that with fixed magnitude for each of the parameters $\sigma_0$, $E$, $\nu$ and $D_f$ the parameter $\chi$ varies only due to changes in the value of $\delta_C$. However, we rely on the analysis of Tang et al.[22] to ensure that this treatment is adequate, that independent variation of $\sigma_0/E$ and $D_f/\delta_C$ is not always necessary, and that the behavior of the pull-off strength conforms to the functional form stated in Eqs. (2) & (3).

**Results for fibril detachment**

Figure 2(a) shows the results obtained for a fibril with a mushroom cap characterized by $\beta = 1.25$ and $\xi = 0.01$. For this geometry, corresponding to a relatively thin stalk and a thin mushroom flange, the configuration depicted in figure 1(c) is used to obtain the results. The average stress $\bar{\sigma}$ applied to the fibril-substrate interface is defined as

$$\bar{\sigma} = \frac{4P}{\pi D_f^2} \qquad (4)$$

is shown as a function of the extension, $\Delta^*$, of the fibril in excess of its elastic extension when fully and rigidly adhered to the rigid substrate. Thus $\Delta^*$, which we term the reduced extension of the fibril, is defined by

$$\Delta^* = \Delta - \frac{P}{k} \qquad (5)$$

where $k$, in this case equal to $1.065 \, \pi E D^2/4L$, is the stiffness of the fibril when it is fully and rigidly attached to the rigid substrate and subjected to an extension $\Delta$. By use of $\Delta^*$, the contribution from the structural extension of the fibril, i.e., its length and stalk diameter dependent response, is largely eliminated from the plotted results.

We first consider a fibril that has $\delta(r) \leq \delta_C$ everywhere on its lower surface, and that, therefore, $c = 0$. At zero applied load its state is located at the origin of figure 2(a), i.e., the applied stress and the reduced extension are both zero, as is the total extension. This condition reflects the fact that, during application of the load prior to attachment and during the process of attachment, no-slip conditions prevailed between the surfaces of the fibril and the substrate. As a result, at zero applied load there are no residual stresses or strains in the fibril and its length is the undeformed value. When a tensile load is first applied, the applied stress increases but the reduced extension remains zero. As a consequence, the location in figure 2(a) that represents the state of the fibril rises vertically up the ordinate from the origin. Even at small loads, separation begins at the edge of the fibril. As the load increases, the location of the maximum separation shifts towards the center of the fibril and does so until the black triangle is reached. When the black triangle is reached, separation initiates at the center of the fibril between the bottom of the fibril and the substrate, but with a cohesive traction, $\sigma_0$, acting in the separated region. When the applied load rises above the level given by the black triangle, the separated segment of the interface will increase in size, in this case forming a circle at the center of the fibril having radius $a$ within which $\delta > 0$. Since $c = 0$ at this stage, the Dugdale zone size parameter $b = a$ (see figure 1(c)). As the applied load is increased, the state of the fibril follows the dark blue curved line that is the upper envelope in figure 2(a). This behavior prevails as long as $c = 0$, and therefore when the cohesive traction, $\sigma_0$, is effective anywhere where there is separation between the bottom surface and the rigid substrate. This implies that $\delta_C$ is very large or unlimited in magnitude. We have repeated this same set of calculations for $c = 0$ using interface strengths of $\sigma_0 = 0.05\ MPa$ and $\sigma_0 = 0.2\ MPa$. In all cases, the resulting normalized pull-off curves are identical to that obtained for $\sigma_0 = 0.1\ MPa$, confirming that the pull-off behavior depends solely on the parameter $\chi$ and not on the individual values of $\sigma_0$ and $\delta_C$, in agreement with the dimensional arguments of Tang et al.[22]

Now consider a situation in figure 2(a) where $\delta_C$ has a finite value with a magnitude that, when tension is applied to the fibril, is eventually reached and exceeded by $\delta$ at the center of the fibril. For a specific value of $\delta_C$, and therefore of $\chi$, the condition $\delta(0) = \delta_C$ is reached at a colored triangle marked on the curved dark blue line in figure 2(a). The value of $\chi$ at which this occurs, and thus the color of the relevant triangle, is determined by $\delta_C$ according to Eq. (3), where the appropriate value of $\chi$ is given in figure 2(a) at the bottom of the line of circles and dashes having the same color as the colored triangle of interest. After $\delta$ at the center of the fibril has exceeded $\delta_C$, the results shown in Figure 2(a) by circles and dashes of a given color specify the load deflection curve for the associated value of $\chi$. Therefore, these results prevail with adhesion conditions in which $c > 0$, with $c$ increasing in magnitude as the colored line consisting of circles and dashes is traversed from top to bottom, and $a$ increasing in magnitude throughout this process as well. Thus, for a given value of $\chi$, the full load deflection curve for the fibril is given by starting at the origin, going up the ordinate to the black triangle, following the upper curved line to the triangle of the relevant color, and then going down the line composed of circles and dashes of the same color. The average stress at the colored triangle is the pull-off strength, $\sigma_C$, for a given value of $\chi$. This conclusion can be made because the load deflection behavior in figure 2(a) is clearly unstable for both load and displacement control at the colored triangles. This instability is also reflected in the post-peak behavior of the extension $\Delta^*$. Immediately after the pull-off point, the curve shows a sharp drop in $\Delta^*$, indicating that the fibril cannot maintain equilibrium under partial detachment. The system responds with rapid unloading. In this case, detachment initiates at the center of the fibril, and during the unstable phase that follows, the contact radius decreases quickly from the center outward.

We emphasize that the curves of a given color are assembled by bringing together finite element results obtained from different meshes, each one with $\sigma_0$ applied over a different value of $c$, and the results unified by having equal values of $\delta_C$. This approach is a convenient one for obtaining accurate and reliable results for the unstable pull-off process.

We note that the pull-off strengths for the nine cases considered in figure 2(a), as indicated by the colored triangles, are all equal to or greater than $0.82\sigma_0$, and are therefore all relatively high. In fact, the highest pull-off strength, for $\chi = 0.05$, is equal to the cohesive strength $\sigma_0$, showing that this case lies within the flaw insensitive regime. Clearly, values of $\chi$ greater than 2.5 for the fibril geometry considered will yield pull-off strengths lower than $0.82\sigma_0$. The location of the black triangle in figure 2(a) suggests that for pull-off strengths lower than $0.78\sigma_0$, detachment will initiate at the edge of the fibril.

Results for a fibril with a thicker stalk and flange ($\beta = 1.05, \xi = 0.05$) are shown in figure 2(b). For this configuration the stiffness of the fibril is $k = 1.076\pi E D^2/4L$. For this geometry, detachment initiates at the edge of the fibril, and the procedure that we have followed for obtaining the results is that associated with figure 1(d). The interpretation of figure 2(b) is analogous to that of figure 2(a), showing similar trends. Notably, pull-off at the colored triangles is also unstable under load control.

The pull-off strengths for the nine cases considered in figure 2(b) range from $0.5\sigma_0$ to $\sigma_0$. For values of $\chi$ less than about 1.1 the pull-off strength will be equal to the cohesive strength, $\sigma_0$, showing that these cases will lie within the flaw insensitive regime. Values of $\chi$ greater than 8 for the fibril geometry under consideration will yield pull-off strengths lower than $0.5\sigma_0$. Also, it is important to note, when figures 2(a) and 2(b) are compared, that for the second case the flaw insensitive regime is reached for higher values of $\chi$, i.e., for lower adhesion energies.

A key feature of both figures 2(a) and 2(b) is that the maximum applied stress, i.e., the pull-off strength under load control, marked by the colored triangles, coincides with the deformation state in which, under rising applied stress, the maximum separation between the lower surface of the fibril and the surface of the rigid substrate first attains the value $\delta_C$. Furthermore, detachment initiates either at the edge of the fibril proceeding inwards from there (for the geometry considered in figure 2(b) and for the geometry considered in figure 2(a) with low strength adhesion) or at the center of the fibril proceeding outwards from there (geometry in figure 2(a) with higher strength adhesion). Unstable detachment then follows under load control in both cases. As a result, the assumptions underlying the deductions of Tang et al.[22] are fulfilled, confirming that the pull-off strength is a function only of $\chi$, $\beta$, $\xi$ and $R/D_f$, as in Eq. (2), and not on individual dimensionless groups that can be formed from the parameters controlling $\chi$. As a consequence, figures 2(a) and 2(b) fully characterize the pull-off strengths for the geometries considered, and no further variation of cohesive parameters is necessary to complete our study of the adhesive strengths of these fibrils, at least within the ranges of parameters considered.

**Results for pull-off strength**

The pull-off strengths for all the geometries considered in this work are shown in figure 3 as a function of $\chi$. The insights of Tang et al.[22] allow us to display our results in terms of the parameters $\chi$, $\beta$ and $\xi$ without the involvement of any other combination of parameters.

Each colored line in figure 3 represents results for fibrils with the same mushroom flange to diameter ratio, i.e., $\beta$, while each symbol (i.e., triangle, square and circle) denotes results for a specific ratio of cap thickness to cap diameter (i.e., $\xi$). For comparison with the results for mushroom fibrils, the result by Fleck et al.[26] for a straight punch subject to interface no-slip conditions at detachment is shown by a dashed black line in figure 3. The Fleck et al. [26] result is obtained for Dugdale cohesion between the fibril and the substrate in the absence of a pre-existing adhesion defect. Because the initial, unloaded state prior to detachment in the Fleck et al.[26] calculations is stress-free, their results for a straight punch are obtained under the same conditions as our results.

As can be seen in figure 3, we can consider the curves to be made of at least two types of segments. The first type of segment (type 1) corresponds to plots similar to that obtained by Fleck et al.[26], i.e., they are like the dashed black line in figure 3. Type 2 segments are less steep and remains relatively close to the line $\sigma = \sigma_0$. All curves also exhibit a segment, denoted the flaw insensitive regime, where the pull-off strength is equal to $\sigma_0$ and is independent of $\chi$. To the right of this segment, each of the curves has a negative slope where the pull-off strength degrades as $\chi$ is increased. Curves of type 1 are associated with geometries such as $\beta = 1.05$ and $\xi = 0.05$, i.e., the geometry with results represented in figure 2(b), while type 2 behavior is observed for the geometry having $\beta = 1.25$ and $\xi = 0.01$, i.e., the geometry with results represented in figure 2(a). The latter curve remains close to the line $\sigma = \sigma_0$ until $\chi = 8000$, beyond which value of $\chi$ a type 1 segment appears. In addition, a small number of the curves shown in figure 3 have a small segment that is neither type 1, nor type 2 nor a flaw insensitive segment.

If we consider the curves, or parts of curves, in figure 3 that correspond to type 1, we see that the following trend arises: the wider the stalk and the thicker the mushroom flange, the closer the curve is to that of Fleck et al.[26]. For this type of curve, if the diameter of the stalk or the thickness of the flange decrease, the strength versus $\chi$ curve moves upwards and rightwards in figure 3, increasing the pull-off strength at a given value of $\chi$. On the other hand, if we consider the type 2 segments of curves, the curve moves upwards in figure 3 through increase of the diameter of the stalk or the thickness of the flange. Therefore, the effect of the relative dimensions of the mushroom flange and the stalk on the two types of strength versus $\chi$ curve is opposite. We will analyze both types of curves separately.

If we consider the curve in figure 3 representing the results for $\beta = 1.25$ and $\xi = 0.01$, i.e., the curve with the larger type 2 segment that is a blue line with a circle on it, the pull-off strength is equal to the cohesive strength, $\sigma_0$, for values of $\chi$ up to 0.06. For larger values of $\chi$, the pull-off strength falls gradually as $\chi$ is increased; however, the decay is so gradual that even for very large values of $\chi$, approaching 8000, the pull-off strength in each case is larger than $0.78\sigma_0$. This situation is consistent with the results in figure 2(a), where we can conclude that the black triangle represents the pull-off strength for $\chi = 8000$. As a consequence, we can expect that the strength versus $\chi$ curve representing these results in figure 3 will have an extensive segment from $\chi = 0.06$ to $\chi = 8000$ in which the pull-off strength lies between $\sigma_0$ and approximately $0.8\sigma_0$. This behavior is also consistent with the results shown in figure 4, depicting, at fibril pull-off for this geometry as functions of radial coordinate, $r$, both the traction, $\sigma$, on the bottom surface of the fibril (continuous lines) and the separation, $\delta$, (dashed lines) between the substrate and the bottom of the fibril. The separation profiles clearly show the movement of the maximum separation point from the edge of the fibril at $\chi = 8000$ to the center of the fibril at $\chi = 2.5$. Over a wide range of values of $\chi$ the distribution of traction varies only slightly, in such a way that the pull-off strength does not vary significantly, even though the separation profile changes noticeably over the same range of $\chi$. Furthermore, it is clear from comparison of Figure 4 with Figure 3 that detachment initiates at the fibril flange edge for $\chi > 8000$ and that the pull-off strength degrades rapidly with increasing $\chi$ beyond $\chi = 8000$ due to such edge detachment initiation; i.e., for $\chi > 8000$ the curve in Figure 3 for this fibril geometry is of type 1.

In contrast, for the strength versus $\chi$ curve in figure 3 representing the results when $\beta = 1.05$ and $\xi = 0.05$, i.e., the type 1 curve that is a yellow line with a triangle on it, the pull-off strength is equal to the cohesive strength, $\sigma_0$, for values of $\chi$ as high as approximately 1. For larger values of $\chi$ the pull-off strength falls quite rapidly as $\chi$ is increased, approaching a value of $0.07\sigma_0$ when $\chi$ is close to 1000 and $0.03\sigma_0$ when $\chi$ is close to 10000. Consistent with this, the traction distribution on the fibril bottom surface at pull-off for this geometry, shown in figure 5 as continuous lines, changes significantly as $\chi$ decreases, leading to a marked modification in the pull-off strength for a relatively small variation of $\chi$. The dashed lines in Figure 5 demonstrate that detachment for this fibril geometry always initiates at the fibril flange edge.

As noted above, figures 4 and 5 show the stress, $\sigma$ (continuous lines), and the separation, $\delta$ (dashed lines), between the fibril surface and the substrate at pull-off for 2 fibril shapes over a range of $\chi$ values. The stress and the separation are shown as functions of radial position on the surface of the fibril and are normalized, respectively, by the cohesive stress, $\sigma_0$, and the critical separation, $\delta_C$. Figure 4, for a fibril with a relatively wide and thin mushroom flange, demonstrates that, in this case, there are Dugdale zones from the center of the fibril almost all the way to its perimeter. In addition, the maximum separation moves from the edge to the center of the fibril as $\chi$ decreases. For high values of $\chi$, the red dashed line indicates that detachment initiates at the edge of the fibril. In this case, separation is already present and highly localized at the perimeter. Since separation is normalized by $\delta_C$, the values shown are exactly 1 where detachment occurs and 0 elsewhere, producing the apparent jump. Additionally, the presence of a ring-shaped crack leads to a stress concentration near the edge, which is reflected in the solid line (normal traction) as a pronounced peak at the perimeter. We note that there are some minor discrepancies in regard to the accuracy of the position of the outer radius of the Dugdale zone when it is compared to the location of the transition from zero to finite separation. However, such discrepancies do not undermine the relative accuracy of the calculations for pull-off strength because it varies only slightly over the range of $\chi$ values utilized in the figure. In figure 5, for a fibril with a relatively narrow and thick mushroom flange, Dugdale zones can be seen at the outer radius of the fibril, with a significant change in their width over the range of $\chi$ considered for the figure. In addition, the maximum separation in this case occurs at the outer radius of the fibril, so that detachment commences there. In this case, there is good numerical consistency between locations of the outer radii of the Dugdale zone and the transitions from zero to finite separation, giving confidence in the accuracy of the calculations of pull-off strength. Figures 4 and 5 together show that fibrils having a beneficial design of a wide and thin mushroom flange experience detachment initiation at the center of the fibril and exhibit only a modest loss of pull-off strength as $\chi$ is increased up to values as high as 8000, whereas poorly designed fibrils, having a narrow and thick mushroom flange, show detachment initiation at the fibril outer radius and a significant decrease in pull-off strength when $\chi$ is increased.

So far we have considered the pull-off strength curves in figure 3 for $\beta = 1.25$ and $\xi = 0.01$, a fibril with a relatively wide and thin mushroom flange that has a type 2 plot over a significant range of $\chi$ values in figure 3, and for $\beta = 1.05$ and $\xi = 0.05$, a fibril with a relatively narrow and thin mushroom flange that has a type 1 plot over a significant range of $\chi$ values in figure 3. As far as the rest of the pull-off curves in figure 3 are concerned, some of them cannot be classified entirely as type 1 or type 2 for the full range of $\chi$ analyzed; i.e., some of the curves show three different segments, with one of these having type 1 characteristic and one having type 2 characteristic. In such cases, one segment of the plot in figure 3, at the smallest values of $\chi$, has a pull-off strength equal to $\sigma_0$, i.e., the flaw insensitive regime, another segment, at intermediate values of $\chi$, is in the type 2 category where the pull-off strength, $\sigma_C$, falls gradually as $\chi$ is increased, and the final segment, at the largest values of $\chi$, is in the type 1 category where the pull-off strength, $\sigma_C$, falls more rapidly as $\chi$ is increased. This feature is present in the case of the geometry with $\beta = 1.15$ and $\xi = 0.01$, represented by a green line with a circle on it in figure 3. The portion of the pull-off strength versus $\chi$ curve corresponding to type 1 can be seen at values of $\chi$ ranging from 10000 to 150, while for $\chi$ values below 13 to the point where the flaw insensitive regime is reached at approximately $\chi = 0.3$, the curve has the type 2 form. Between $\chi = 150$ and $\chi = 13$ the curve is dissimilar from both type 1 segments and type 2 segments. In this case, moving the point of maximum separation from the edge to the center of the fibril results in a larger variation in the stress distribution than in the geometry with $\beta = 1.25$ and $\xi = 0.01$.

The transition between the type 1 and type 2 curves can be better appreciated if we analyze the fibril with $\beta = 1.25$ and $\xi = 0.025$, represented by a blue curve with a square on it in figure 3. The portion of the pull-off strength versus $\chi$ curve corresponding to type 1 can be seen at values of $\chi$ ranging from 10000 to 58, while for $\chi$ values below 20 to the point where the flaw insensitive regime is reached at approximately $\chi = 0.25$, the curve has the type 2 form. Separation profiles at the base of this fibril, for four different $\chi$ values bridging the transition from type 1 curve to type 2, are shown in figure 6 as dashed lines. Note that each case illustrated in figure 6 has some separation occurring at the very edge of the fibril, albeit in a very narrow annulus. The highest $\chi$ value illustrated, $\chi = 61$, places itself just inside the type 1 segment of the pull-off strength versus $\chi$ curve, as can be seen in figure 3. The separation profile for this case, shown as the red dashed line in figure 6, clearly indicates that detachment initiates at the edge of the fibril as the separation elsewhere than in the perimeter annulus is zero. The profile for $\chi = 60$, represented in figure 6 by the green dashed line and associated, barely, with the type 1 segment of the pull-off strength curve in figure 3, shows separation at two different locations, at the edge of the fibril and at its center. However, the maximum separation is located at the outer radius of the fibril, i.e., at $r = D_f/2$, indicating that detachment initiates there. At $\chi = 58$, where the pull-off strength curve in figure 3 transitions from type 1 to type 2, the separations at the center of the fibril and at its outer radius are equal, as can be seen in the separation profile represented by the yellow dashed line in figure 6. Nevertheless, separation extends significantly from the center of the fibril, while edge separation remains highly localized. In this case, we deduce that detachment is initiated simultaneously at the fibril center and at its outer radius. For smaller values of $\chi$, where the pull-off strength curve in figure 3 is in the type 2 category, the maximum separation location is at the center of the fibril, as can be seen in the profile for $\chi = 7$, represented by a blue dashed line in figure 6. In this case, detachment is initiated at the center of the fibril. These observations suggest that the curve type is governed by the position of the maximum separation, either at the center or the edge of the fibril. If separation initiates at the outer radius of the fibril, the pull-off strength versus $\chi$ plot for that case will be type 1. If separation initiates at the center of the fibril, the pull-off strength versus $\chi$ plot for that case will be type 2 or the fibril will be in the flaw insensitive regime.

Figure 6 also shows the normal stress, $\sigma$, at the interface between the fibril and the substrate for the specified values of $\chi$ for the case of the geometry with $\beta = 1.25$ and $\xi = 0.025$, represented by a blue line with a square on it in figure 3. The stress distributions show the two main characteristics previously discussed for these fibrils: a Dugdale zone extending from the center of the fibril, corresponding to a separation zone there, and a stress concentration at the fibril outer radius corresponding to separation at that location. A notable feature is that all these stress profiles are quite similar. From this, we conclude that the pull-off strength versus $\chi$ behavior is tied closely to how the separation develops, but with the stress distribution not varying greatly in the transition that occurs from detachment initiation at the center of the fibril to detachment initiation at its outer radius.

**Pull-off strength of fibrils with an initial adhesion defect**

An inference from the results presented in figures 3 to 6 and described in the previous section is that it is desirable to have detachment of a fibril from a substrate initiate at the center of a fibril as that enables the achievement of a relatively high pull-off strength. We established that when detachment is initiated at the center of the fibril, the plot of pull-off strength versus $\chi$ is then of type 2, enabling identification of such cases by inspection of figure 3. We note that figure 3 shows that such cases lead to a pull-off strength that is near 80% of the fibril-substrate cohesive strength, $\sigma_0$, and therefore they have very good adhesion. However, these results were carried out in the absence of adhesion defects that will degrade the pull-off strength of the fibril. Such adhesion defects are segments of the interface where the cohesive stress is very low or zero even when the fibril is otherwise adhered to the substrate. We model these adhesion defects as regions where the cohesive stress between the fibril and the substrate is zero. One possibility leading to this situation is that there is a depression in the surface of the fibril so that the distance between the fibril surface and the substrate exceeds the critical value, $\delta_C$.

To investigate the effect of such adhesion defects we place one at the interface between the fibril and the substrate at the center of the fibril. Elsewhere, the cohesive stress between the fibril and the substrate is $\sigma_0$ as before. These defects are therefore represented by the configurations shown in figures 1(b) and 1(c), with $2c$ there being the initial diameter of the defect. Such a defect will have little effect on the pull-off strength of fibrils that experience detachment initiation at the outer radius of the fibril, and thus we assess the effect of adhesion defects only for those fibril designs that would, in the absence of a defect, initiate detachment at the center of the fibril over a significant $\chi$ range. Therefore, we compute and present results for the three geometries with more extensive type 2 curves: geometries with $\xi = 0.01$ and $\beta$ equal to 1.15, 1.2 and 1.25, i.e., curves represented by a green line, a red line and a blue line respectively with a circle on them in figure 3. Note that these geometries correspond to the thinnest and the three widest mushroom flanges. In this study, we have focused on defects located at the center of the fibril–substrate interface. This choice is motivated by the fact that, in the geometries analyzed, the maximum tensile traction typically occurs at the center. Therefore, a defect located there is the most likely to initiate premature detachment and has the greatest potential impact on pull-off strength. While defects may appear elsewhere in practice, they are less likely to be activated under the loading conditions considered.

The procedure for computing the effect of the adhesion defect on the pull-off strength of the fibril is identical to the methodology described above, except that calculations for cases are excluded where $2c$ in figures 1(b) and 1(c) is less than the initial diameter of the adhesion defect. For each fibril geometry studied, initial defect diameters ranging from $2c/D_f = 0.0005$ to $2c/D_f = 0.25$ are assessed.

Figure 7 shows the pull-off strength for a fibril with a mushroom flange defined by $\beta = 1.25$ and $\xi = 0.01$, i.e., one with a relatively wide and thin flange and that has a pull-off strength versus $\chi$ curve in the absence of an adhesion defect in figure 3 shown in blue with the circular symbol on it. The latter curve is reproduced in figure 7 as that marked $2c/D_f = 0$. Each continuous curve in figure 7 represents the strength of the fibril with an initial defect size with the value of $2c/D_f$ marked in the plot, where $2c$ is the initial diameter of the adhesion defect. The fibril pull-off strength is defined as the total force at pull-off divided by the total area of the mushroom flange surface, as in Eq. (2).

As can be seen from figure 7, at larger values of $\chi$ the presence of a defect in the interface at the center of the fibril causes the pull-off strength to decrease notably. The reduction observed is a function of both the size of the defect and the parameter $\chi$, with the pull-off strength decreasing both with increasing $\chi$ and increasing defect size. At the smallest $\chi$ values considered, the strength reduction due to the defect is minimal and mainly attributable to the reduction in effective contact area. For example, when $2c/D_f = 0.25$, the strength reduction is 6.25%, which is due to the area reduction over which the cohesive stress $\sigma_0$ is sustained. The limit of this behavior, in which the defect does not significantly alter the resistance, is clearly dependent on the size of the defect, and can persist to values of $\chi$ near 600 for the smallest defect. However, for values of $\chi$ above this limit, for each defect size considered, the pull-off strength drops steeply away from that of the fibril without the defect, showing that in this range of $\chi$ the adhesion defect has a significant effect on the pull-off strength of the fibril.

In the range where the pull-off strength with the defect present is much less than the strength in the absence of the defect, i.e., at very high values of $\chi$, the Dugdale zone is much smaller than the fibril flange radius, and thus the defect is functioning as a crack in a linear elastic system with a very small nonlinear zone of material at its tip. In this regime, linear elastic fracture mechanics will prevail, although the no-slip condition at the interface can cause the behavior of the defect to be mixed mode in fracture mechanics terms. However, for the very shortest cracks, the mode mixity is almost pure Mode I when the substrate is rigid, and in the limit of a vanishingly small crack for such a substrate, Kassir and Bregman [33] show that the penny shaped crack at the interface is purely Mode I. In addition, for the no slip interface at a rigid substrate, Kassir and Bregman [33] identified the Mode I stress intensity factor for an interface crack that is small compared to the component diameter to be the same as that at a slipping interface. Thus, for such a very small penny-shaped crack of radius $c$ the Mode I stress intensity factor is

$$K_I = \frac{8P}{\pi^2 D^2}\sqrt{\pi c} \tag{6}$$

and the Mode II stress intensity factor is zero. Accordingly, the energy release rate is given by

$$\mathcal{G} = \frac{(1-\nu^2)K_I^2}{2E} = \frac{32(1-\nu^2)P^2 c}{\pi^3 E D^4} \tag{7}$$

We equate this to the work of adhesion, and predict the pull-off strength for a mushroom fibril with a very small adhesion defect at the center to be

$$\frac{\sigma_C}{\sigma_0} = \frac{1}{\beta^2} \sqrt{\frac{D_f/2c}{2\chi}} \qquad (8)$$

The expression (8) has been plotted with dashed lines in figure 7, each dashed line representing an initial defect size, $2c/D_f$. The continuous curves in the same colors represent the corresponding pull-off strength obtained numerically with the Dugdale zone present. As shown, for the whole range of defects considered, both the numerical results (solid lines) and LEFM predictions (dashed lines) yield nearly identical pull-off strengths at high $\chi$ values, showing that in those cases, when the size of the Dugdale zone is very small, the defect behaves like a crack in a linear elastic system. It is notable that, even at lower values of $\chi$, the predictions obtained from linear elastic fracture mechanics, shown as dashed lines in figure 7, other than in the defect insensitive regime, are relatively accurate despite the presence of Dugdale zones of significant extent.

Pull-off strength results with different defect sizes have been similarly obtained for the other two geometries under consideration. Figure 8 shows the pull-off strength in the presence of defects, together with the predictions of equation (8), for a fibril with $\beta = 1.2$ and $\xi = 0.01$, while figure 9 does the same for a fibril with $\beta = 1.15$ and $\xi = 0.01$. For each of those geometries, the results show that the effect of an adhesion defect in the interface at the center of the fibril is similar to that shown in figure 7 for the geometry having $\beta = 1.25$ and $\xi = 0.01$. That is, there is a decrease in strength with increasing defect size and with increasing $\chi$, and in each case there are two clearly defined parts in the plot of pull-off strength versus $\chi$. With increasing $\chi$ these are a defect insensitive segment where the strength is simply decreased by the reduction in area that can sustain cohesive stress, and a segment at the largest values of $\chi$ where the strength drops steeply below the strength of the fibril without the defect and the defect is acting as a crack in a linear elastic system.

It should be noted that in figure 8 the curve corresponding to the smallest defect has not been represented, as it would coincide with the curve without defect. Therefore, a defect of such small size ($2c = 5 \times 10^{-4} D_f$) does not alter the pull-off strength when compared to the defect free case. That is, for $\chi < 360$ detachment initiates at the center of the fibril, but for higher values of $\chi$ the presence of such a small defect at the center of the fibril does not prevent separation initiating at the edge. A similar situation arises in figure 9, where the pull-off strength curves are plotted for the fibril with $\beta = 1.2$ and $\xi = 0.01$ with different sizes of central defects. In this case, the two smallest defects, with sizes $2c = 5 \times 10^{-4} D_f$ and $2c = 1.25 \times 10^{-3} D_f$, do not alter the pull-off behavior of the fibril compared to that without the defect. In this case, separation starts at the edge of the flange for $\chi > 160$.

**Discussion**

We have analyzed the behavior of mushroom shaped fibrils adhering to a rigid substrate and obtained predictions of their pull-off strength under load control, where the Dugdale model of cohesion is assumed to provide the adhesion between the fibril and the rigid substrate. We note that the pull-off strength shown in figure 3 could be fitted to curves to represent our results. Such an approach has been used by Aksak et al[24], who found that their results for pull-off strength, $\sigma_C$, for a fibril with a conical mushroom flange could be fitted according to

$$\frac{\sigma_C}{\sigma_0} = \begin{cases} B_e \left[ \hat{\beta} \chi \right]^{-\alpha} & (a) \\ B_c \left[ \hat{\beta} \chi \right]^{-0.5} + \Gamma_c & (b) \end{cases} \qquad (9)$$

where $B_e$ and $B_c$ are geometry dependent constants, $\Gamma_c$, also geometrically dependent, is a saturation strength at high values of $\chi$, and $\alpha$ is an exponent dependent on the contact angle at the edge of the fibril. The parameter $\hat{\beta}$ is the ratio of the tip diameter of the conical mushroom flange to the stalk diameter. The expression in Eq. (9a) applies to the cases where separation initiates from the fibril edge, and the one in Eq. (9b) to the case where separation initiates from the center of the fibril tip. Eq. (9a) can be interpreted as describing type 1 curves in figure 3, while Eq. (9b) corresponds to curves of type 2 behavior. We have not attempted to check that such fits are accurate for our results in figure 3. In any case, the use of the fits in Eq. (9) will lack some detail, as there are multiple segments to many of the curves that would not be captured by this equation. Furthermore, it is by no means clear that an exponent of -0.5 will be universally correct for the type 2 results in figure 3. For all these reasons, we have not attempted to provide fits to the curves there, whether according to Eq. (9), or involving any other elementary functions.

Nevertheless, the curves provide important information about the relation between mushroom flange shape and pull-off strength. For all the geometries considered in this work, it is possible to distinguish at least two different segments in the strength curve: for small values of $\chi$ all geometries in figure 3 present a horizontal segment corresponding to the flaw insensitive regime, and for large values of $\chi$ all geometries present a segment that descends steeply versus $\chi$ and that corresponds to detachment initiating from the edge of the flange. We term the latter a type 1 curve. Between these two segments, some of the geometries have curve segments differing from type 1. To quantify the adhesion performance of the different geometries considered, we define two threshold values of $\chi$: $\chi_{c1}$, below which the flaw independent regime is reached with the pull-off strength equal to the intrinsic adhesive strength independent of tip shape[10, 22] and $\chi_{c2}$, above which the strength decreases steeply with $\chi$ in a similar way as occurs for the punch geometry, i.e., a type 1 curve. In between these 2 values of $\chi$ the pull-off strength falls gradually as $\chi$ is increased. Given this insight, we deduce that the best fibril geometry will be one with both $\chi_{c1}$ and $\chi_{c2}$ as high as possible. As we have already noted, the effect of the relative dimension of the mushroom flange and the stalk on these two limits is opposite, and geometries with high values of $\chi_{c1}$ tend to have low values of $\chi_{c2}$.

For $\chi < 3$, all geometries exhibit pull-off strength higher than $0.8\,\sigma_0$, and some geometries have pull-off strength equal to $\sigma_0$ for this range of $\chi$. In this range of $\chi$, in general, geometries with narrow, thick flanges have higher pull-off strengths. However, for $\chi < 3$ some of these geometries exhibit a segment of curve that is type 1, i.e., $\chi_{c2} < 3$, so that the pull-off strength decreases very rapidly as $\chi$ increases. The geometries that maintain high values of pull-off strength to higher values of $\chi$, (i.e., geometries with higher values of $\chi_{c2}$) are those with the widest, thinnest flange ($\xi = 0.01$ and $\beta = 1.25, 1.2$ and $1.15$). It can be seen that at least for $20 \leq \chi \leq 150$, these three geometries exhibit high pull-off strength with values superior to $0.78\,\sigma_0$, much higher than that presented by other geometries, especially the punch. The geometry with $\beta = 1.25$ and $\xi = 0.01$, having the widest, thinnest flange, maintains good pull-off strength up to values of $\chi$ greater than 8000.

The high pull-off strength of mushroom-shaped fibrils with wide, thin flanges compared with that of other geometries identified in our computational simulations agrees with the experimental results obtained by Del Campo et al.[14] When the pull-off force of a mushroom ($\beta = 1.25$) and a punch were compared, both having the same stalk diameter, $D = 20\,\mu m$, they found that the adhesion force for the mushroom is on the order of 20 times higher. Taking the material properties given by the same authors for a similar soft PDMS material[34] ($\sigma_0 =$ 1 MPa; $E = 1.43$ MPa; $\nu = 0.5$; $W_{adh} = 0.068$ J/m$^2$), from equation (1) we obtain $\chi$ values of 30.7 and 24.6 respectively for the mushroom and the punch fibrils investigated by Del Campo et al.[14] For these values of $\chi$, the pull-off strengths from figure 3 are 0.8 $\sigma_0$ for a mushroom with $\beta = 1.25$ and $\xi = 0.01$, and 0.21 $\sigma_0$ for the punch, implying pull-off forces of 393 μN and 66 μN respectively for the mushroom and punch fibrils investigated by Del Campo et al.[14] This corresponds to a pull-off load six times higher for the mushroom than for the punch, significantly lower than the factor of 20 reported by Del Campo et al.[14] However, their measurements were performed using a hemispherical glass surface, a geometry that likely enhanced the effectiveness of the mushroom tip due to conformal contact[21, 35-36].

For each geometry, its plot in figure 3 can be considered a design curve, e.g., allowing selection of an optimum fibril size for a given material. Optimal behavior is achieved if the fibril pull-off strength equals $\sigma_0$, the interface adhesion strength, so that detachment occurs in the flaw insensitive regime. Therefore, to achieve optimal pull-off strength, the design constraint is $\chi \leq \chi_{c1}$, a condition that can only be met for a given material by use of a fibril with a relatively small diameter tip. The resulting dimensional constraint is challenging, with the requirement for a sufficiently small fibril tip being easiest to meet in the case of one having a relatively narrow flange that is relatively thick; figure 3 shows that, of those we investigated, the fibril with $\beta = 1.15$ and $\xi = 0.05$ has the largest value of $\chi_{c1}$, in this case equal to 2.84. Even in this more favorable case, the maximum tip diameter permitted for the flaw insensitive regime is 2.31 μm for the material used for experiments by Del Campo et al.[14] This implies a stalk diameter of 2 μm for this mushroom shaped fibril. Conditions are typically more demanding if the fibril material has stronger adhesion. For example, for keratin ($\sigma_0 = 20$ MPa; $E = 1$ GPa; $\nu = 0.5$; $W_{adh} = 0.01$ J/m$^2$)[13], the required tip size is approximately 4 times smaller.

This stringent size constraint is relaxed considerably if we use as a design criterion that pull-off is initiated at the center of the fibril, implying that a curve of type 2 predicts the pull-off strength in figure 3, i.e., $\chi \leq \chi_{c2}$. In this case, the pull-off strength is not $\sigma_0$, the interface adhesion strength, but, as can be seen in figure 3, the resulting pull-off strength remains reasonably high, never falling below 0.78 $\sigma_0$. The most favorable shape is the fibril with $\beta = 1.25$ and $\xi = 0.01$, i.e., that with the widest, thinnest flange, having $\chi_{c2} = 8000$. In this case, for the material used in experiments by Del Campo et al.[14], the tip diameter can be as large as 6.52 mm with a stalk diameter that is 5.21 mm and, at that size, the pull-off strength is predicted to be 0.78 MPa. For keratin, a mushroom fibril of this shape can have a tip diameter as large as 1.68 mm and, at that size, the pull-off strength is predicted to be 15.6 MPa.

We now address the consequences of there being an adhesion defect at the center of the fibril tip, where figure 7 illustrates its effect on pull-off strength for a fibril with $\beta = 1.25$ and $\xi = 0.01$. If we use the PDMS investigated by Del Campo et al.[14] and design a mushroom fibril with tip diameter 6.52 mm, stalk diameter 5.21 mm and flange thickness 65.2 μm, i.e., $\beta = 1.25, \xi = 0.01, \chi = 8000$, in the absence of an adhesion defect it is predicted to have a pull-off strength equal to 0.78 $\sigma_0 = 0.78$ MPa, as noted above. However, even a very small defect having $2c = 0.0005 \, D_f = 3.26$ μm reduces its pull-off strength to 0.27 $\sigma_0 = 0.27$ MPa (a 65% decrease). A defect ten times larger results in a 96% reduction in pull-off strength (to 0.03 $\sigma_0 = 30$ kPa). For lower $\chi$ values, the effect of adhesion defects of these sizes in the fibril in question is less pronounced: 14% and 90% for $\chi = 1000$, and 1% and 70% for $\chi = 100$. It follows that a more robust design for a fibril made from the PDMS investigated by Del Campo et al.[14] is one that has a smaller value of $\chi$, i.e., a smaller fibril. For example, a fibril of this material having tip diameter 407 μm, stalk diameter 325.6 μm and flange thickness 4.07 μm has $\beta = 1.25$, $\xi = 0.01$ and $\chi = 500$. Figure 7 then shows that a central defect of size $2c = 0.0005 \, D_f = 0.2035$ μm has negligible effect on the fibril's pull-off strength as it remains essentially the same as that of the defect free fibril. On the other hand, a central defect of size $2c = 0.005 \, D_f = 2.035$ μm in this fibril reduces its pull-off strength from the defect free value of 0.78 $\sigma_0 = 0.78$ MPa to 0.27 $\sigma_0 = 0.27$ MPa. Therefore, a design strategy using this fibril geometry ($\beta = 1.25$, $\xi = 0.01$) for materials similar to the PDMS studied by Del Campo et al.[14] is to estimate the maximum expected adhesion defect and to choose a size for the fibril that leads to a value of $\chi$ at which, according to figure 7, the largest likely adhesion defect causes a negligible reduction in the pull-off strength compared to that for an defect free fibril.

The decrease in pull-off strength of a cylindrical punch produced by an adhesion defect has been analyzed by Khaderi et al[37]. Their results, together with figure 7 of this study, allow us to analyze the effect of an adhesion defect on the fibrils used by Del Campo et al[14] in their experiments. We return to the mushroom and punch fibrils previously analyzed in this section, to which correspond values of $\chi$ equal to 30.7 and 24.6, respectively. Let us assume now that there is an initial defect of size 5 μm in both specimens, this defect being located at the center of the fibril tip for the mushroom and at the perimeter for the punch. Figure 7 shows that in this case the strength of the mushroom fibril is reduced to $0.48\,\sigma_0 = 0.48$ MPa, implying a pull-off force of 210 µN. Following Khaderi et al al[37], a punch with $\chi = 24.6$ has its strength reduced to $0.17\,\sigma_0 = 0.17$ MPa by a perimeter defect of 5 µm, giving a pull-off force of 54 µN. That is, the reduction in pull-off force is 46% in mushroom and 17% in punch. Therefore, the negative effect of an initial adhesion defect is greater for the mushroom than for the punch. This effect can be attributed to the fact that mushroom-shaped fibrils are specifically designed to promote central detachment, where stress is distributed more uniformly across the interface. When a defect is located precisely at the center—where adhesion is most critical for such geometries—it directly undermines the mechanism that enables their superior performance. In contrast, punch-shaped fibrils tend to initiate detachment at the edge, making them less sensitive to central defects and therefore less affected overall.

The presence of defects in the central zone of the interface between the fibril tip and the substrate, acting as the primary mechanism for separation in mushroom fibrils, has been experimentally demonstrated [30]. Building upon this experimental evidence, Booth et al.[31] conducted an analysis to assess the strength of a fibril array by considering a statistical distribution of defects on the contact surfaces of individual fibrils leading to a distribution of fibril pull-off strengths within the array. Figures 7, 8, and 9 illustrate the correlation between the strength of each individual fibril and defect size, a relationship that is needed for predicting the effective adhesion strength of the fibrillar array using this statistical model. In addition, the pull-off strength of a mushroom fibril with a central defect can be expressed in functional form as

$$\frac{\sigma_C}{\sigma_0} = f\left(\chi, \frac{h}{D_f}, \frac{D_f}{D}, \frac{R}{D_f}, \frac{2c}{D_f}\right) \qquad (10)$$

extending the Tang et al.[22] analysis to encompass the pull-off strength of a fibril with an adhesion defect on the interface, thereby enabling the statistical prediction of the adhesion strength of a fibrillar array.

**Conclusions**

This study presents a computational analysis of the pull-off strength of mushroom-shaped fibrils adhering to rigid substrates, considering variations in geometry and the presence of adhesion defects. The results confirm that fibrils with a wide, thin mushroom cap exhibit higher pull-off strength compared to those with narrower, thicker caps. This enhancement is attributed to the reduction of stress concentrations at the fibril edge and the promotion of detachment initiation at the fibril center.

Importantly, our results show that for all the fibril geometries studied, detachment can initiate from the fibril edge under specific conditions, regardless of the design. However, detachment from the center does not occur in all geometries, regardless of the conditions. This implies that only certain fibril geometries can fully exploit the adhesion-enhancing mechanisms associated with central detachment. Additionally, we have obtained the full pull-off curves, demonstrating that the separation process is unstable whether it initiates from the fibril center or from its edge. This confirms that, under load control, detachment occurs abruptly once the critical separation is reached.

Our results align with previous experimental studies, demonstrating that optimized mushroom fibril geometries can significantly enhance adhesion performance. Furthermore, we show that the presence of adhesion defects at the fibril center can reduce pull-off strength significantly, particularly for high values of the dimensionless parameter $\chi$. However, selecting fibril dimensions appropriately can mitigate the negative impact of these defects.

The insights gained from this work contribute to the design of advanced bio-inspired adhesives and structured surfaces with improved adhesion properties. Future research should examine additional factors such as substrate compliance, viscoelasticity, and environmental conditions to further refine the predictive models for fibril adhesion behavior.

**Acknowledgements**

CB and CR gratefully acknowledge financial support from the Spanish Ministry of Education in the form of their Senior Faculty mobility grants, which enabled them to undertake research at UCSB. EM-P acknowledges financial support from the People Programme (Marie Curie Actions) of the European Union's Seventh Framework Programme (FP7/2007-2013) under REA Grant agreement no. 609405 (COFUNDPostdocDTU). RMM acknowledges support from an Alexander von Humboldt Foundation Alumni Award.

**Appendix A**

To accurately capture all aspects of the adhesion and detachment processes by the finite element method, the elements at the fibril-substrate interface should be small enough to accurately characterize the behavior of the Dugdale cohesive zone, even at high values of $\chi$. To enable the suitable choice of element size, the extent of the Dugdale zone can be determined using the strip-yield approach[27], in this case for a penny-shaped crack in the interface.

The stress intensity factor for a small, penny-shaped crack of diameter $2c$ subject to a uniform, tensile, interface stress, $\sigma_I$, and a Dugdale cohesive stress, $\sigma_0$, acting at the crack perimeter on an annulus of width $b$ can be obtained as the superposition of the stress intensity factors for the three states shown in figure A1 to obtain a zero stress intensity factor at radius $c + b$. The result establishes a relationship among the interface stress, the cohesive stress, the crack length, $c$, and the Dugdale zone extent, $b$, as

$$\frac{\sigma_I}{\sigma_0} = \frac{\sqrt{(c + b)^2 - c^2}}{c + b} \tag{A1}$$

On the other hand, at high $\chi$ values, the Dugdale zone is significantly smaller than the fibril flange radius, and the system behaves as one with a crack in a linear elastic system with a small cohesive zone at its tip. According to linear fracture mechanics, the pull-off strength for a mushroom fibril with a small adhesion defect is given by

$$\frac{\sigma_I}{\sigma_0} = \sqrt{\frac{D_f}{4c\chi}} \tag{A2}$$

For large values of $\chi$, the results in Eq. (A1) & (A2) are equal to leading order. Therefore, we equate these formulae and obtain an expression for $b$ in the form

$$\frac{b}{c} = \frac{1}{\sqrt{1 - \dfrac{D_f}{4c\chi}}} - 1 \tag{A3}$$

To properly model the effect of the Dugdale cohesive zone, at least 5 finite elements should be present along its length. Using a mesh with element size $L_{ele} = (D_f/2)/210000$, we deduce that our calculations are valid for $\chi$ values up to 10000 for all crack lengths considered in this study as there will always be more than 5 elements along any cohesive zone that develops in the simulations..

Details of a typical mesh are illustrated in figure A2. Along the interface with the substrate there are three rows of four-noded axisymmetric elements with dimensions $L_{ele} \times L_{ele}$. The element size increases significantly with distance from the substrate interface, ensuring that the total number of elements in the model is acceptable for computational purposes.

**Figures**

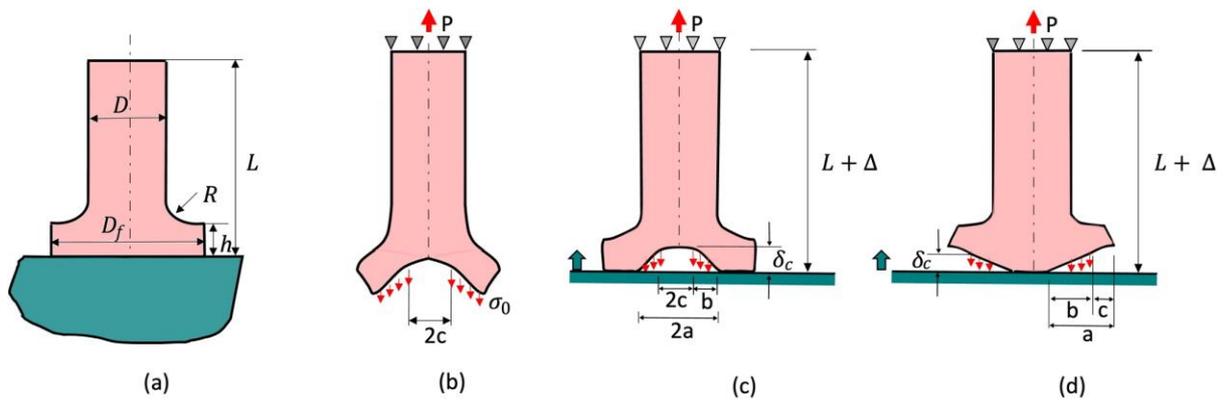

Figure 1. Deformation process for an initially stress free fibril. (a) Geometry of the fibrils; (b) initial deformation state; (c) gradual compression of the mushroom end and development of the Dugdale zone at the center of the fibril-substrate interface and (d) at the perimeter of the fibril mushroom tip.

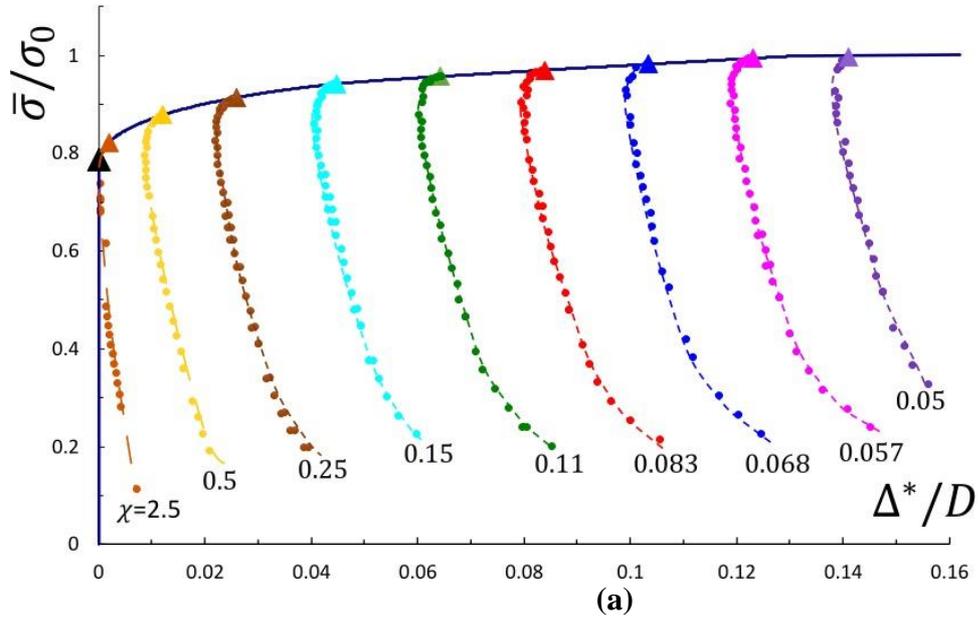

**(a)**

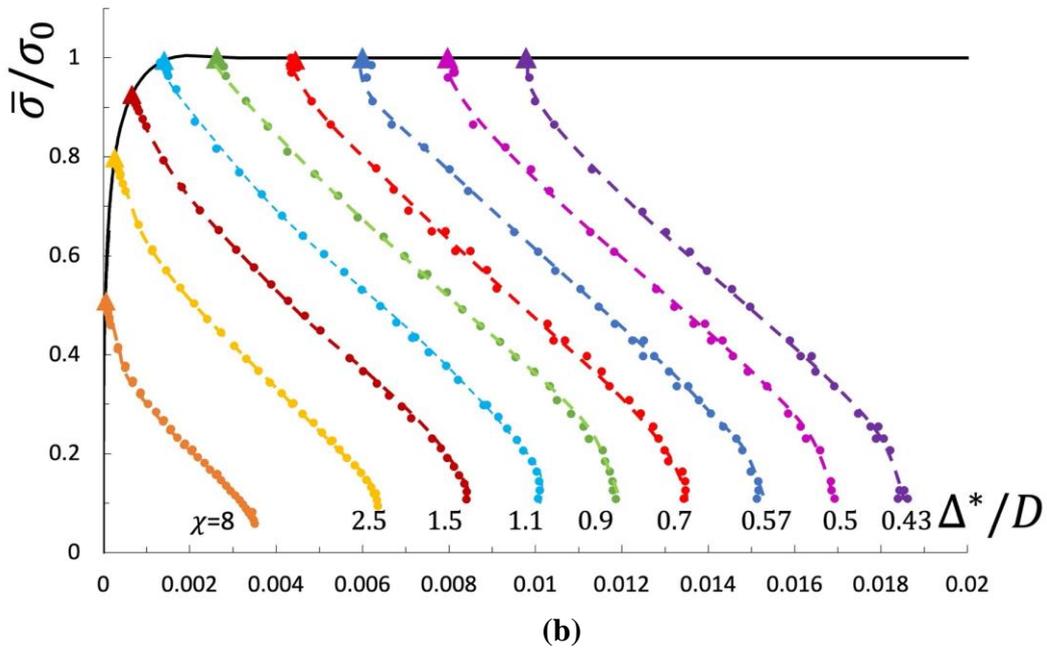

**(b)**

Figure 2. Load-extension behavior for two mushroom fibril geometries under load control: (a) $\beta = 1.25$ and $\xi = 0.01$, where detachment initiated at the center; (b) $\beta = 1.05$ and $\xi = 0.05$, where detachment initiated at the edge. Each coloured triangle denotes the pull-off point for a given value of χ. These curves illustrate the unstable nature of the detachment process and help identify the geometries capable of achieving high pull-off strength through central detachment.

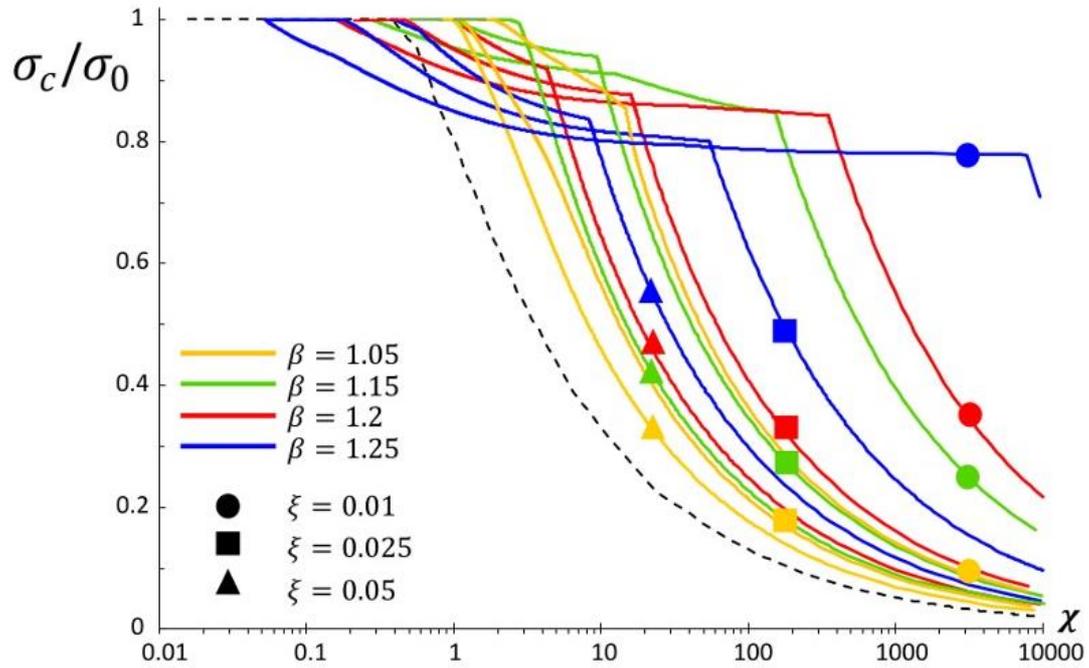

Figure 3. Normalized pull-off strength versus χ for different mushroom fibril geometries. Each curve corresponds to a ratio of flange to stalk diameter ($\beta$), and each symbol type corresponds to a specific flange thickness ($\xi$). The dashed line represents the performance of a punch geometry for comparison[26]. This figure provides a design map to optimize fibril geometry for desired adhesion performance, highlighting the regimes of flaw insensitivity, gradual degradation, and sharp strength loss.

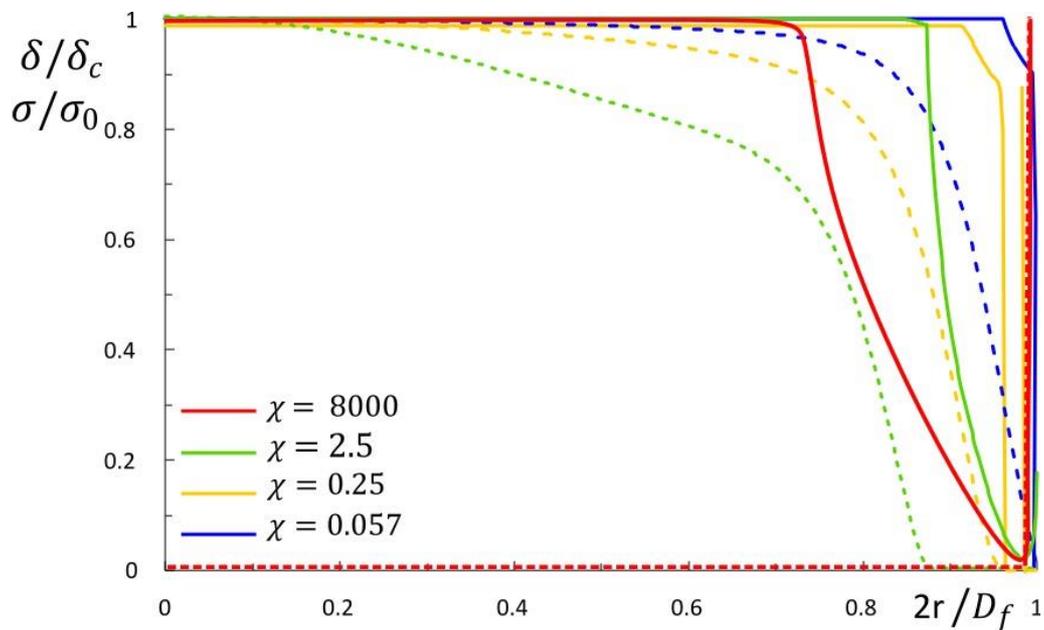

Figure 4. Normal traction and interface separation along the fibril–substrate interface at the moment of pull-off, for a mushroom fibril with $\beta = 1.25$ and $\xi = 0.01$. Results are shown for multiple values of the parameter $\chi$. Solid lines represent the normal traction; dashed lines represent the separation.

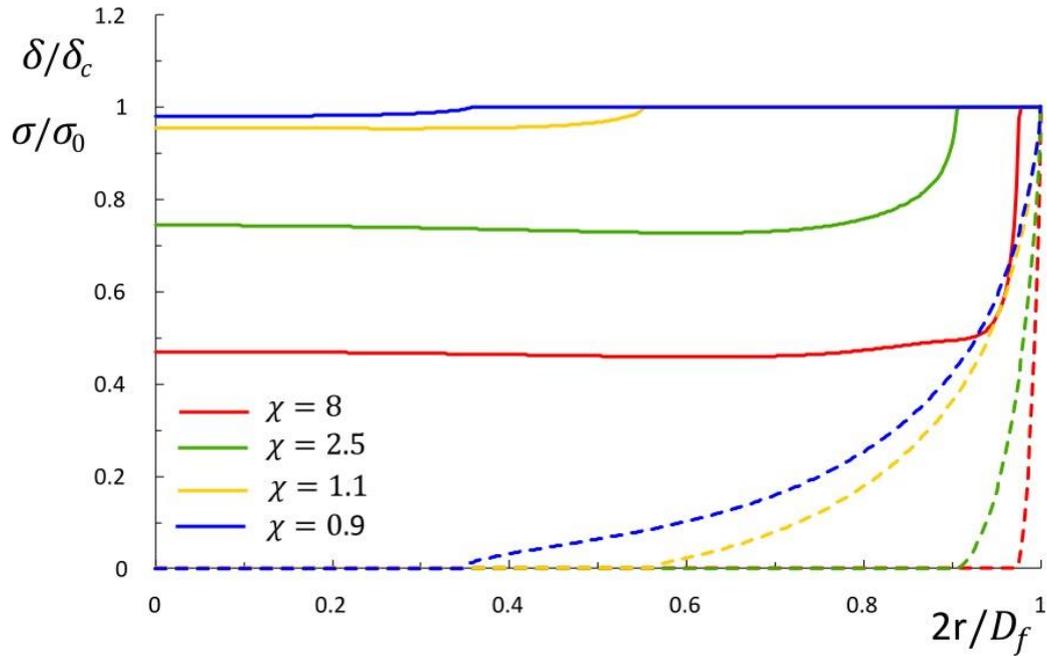

Figure 5. Normal traction and interface separation along the fibril–substrate interface at the moment of pull-off, for a mushroom fibril with $\beta = 1.05$ and $\xi = 0.05$. Results are shown for multiple values of the parameter $\chi$. Solid lines represent the normal traction; dashed lines represent the separation.

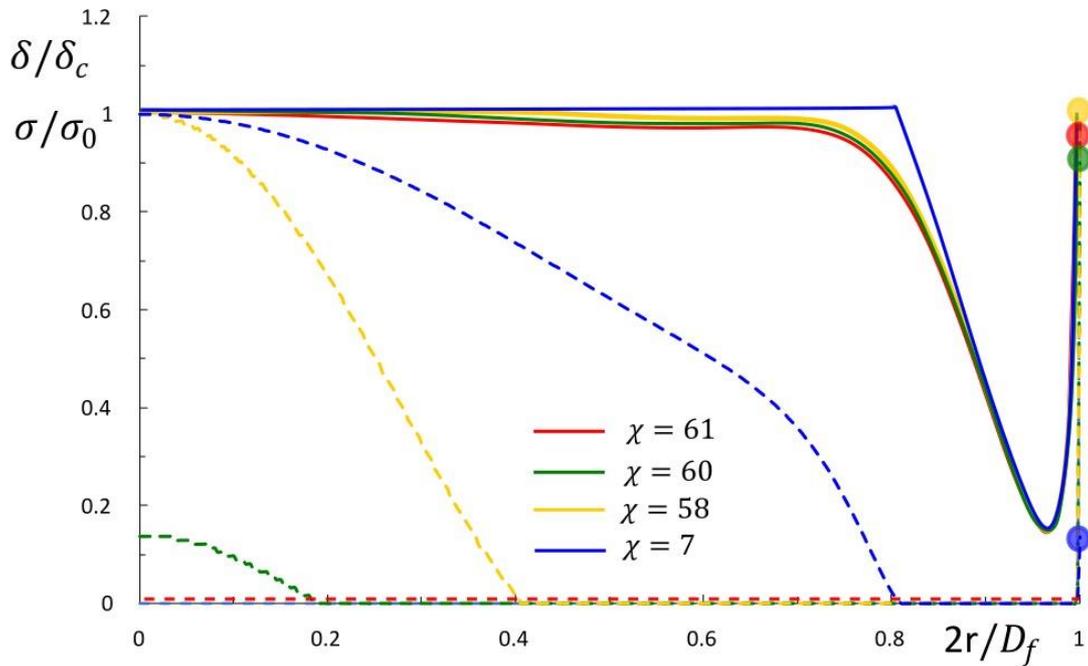

Figure 6. Normal traction and interface separation along the fibril–substrate interface at the moment of pull-off, for a mushroom fibril with $\beta = 1.25$ and $\xi = 0.025$. Results are shown for multiple values of the parameter $\chi$. Solid lines represent the normal traction; dashed lines represent the separation.

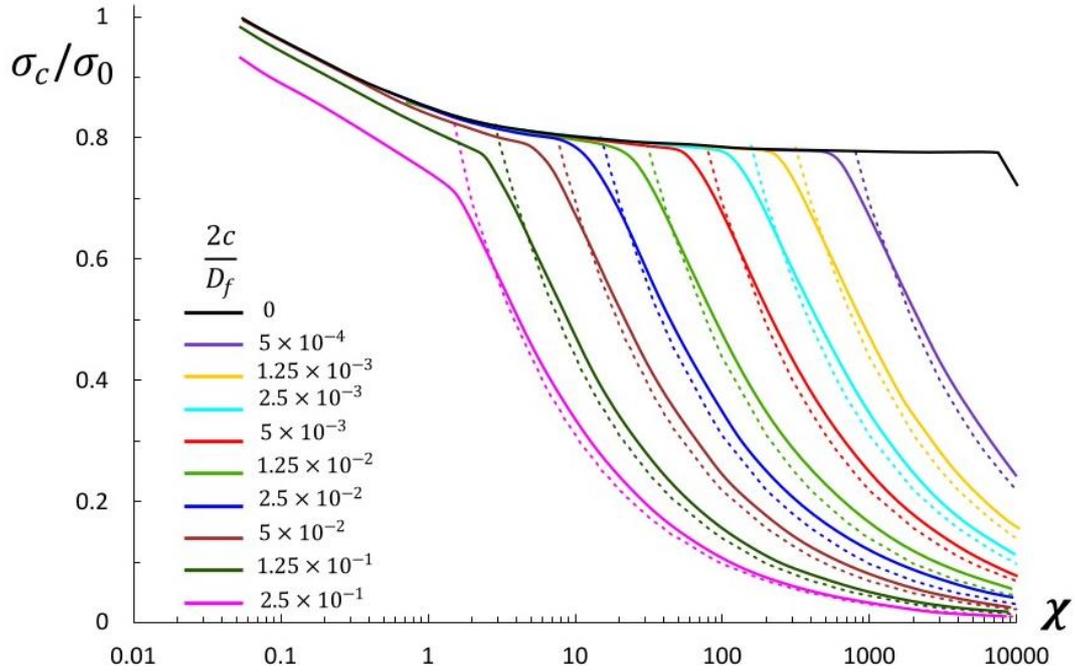

Figure 7. Effect of a central adhesion defect on the pull-off strength of a mushroom fibril with $\beta = 1.25$ and $\xi = 0.01$. Each curve represents a different defect size, $2c$. Dashed lines show linear elastic fracture mechanics (LEFM) predictions for comparison. This figure quantifies the sensitivity of high-performance fibrils to central defects and underscores the need to consider defect tolerance in adhesive surface design.

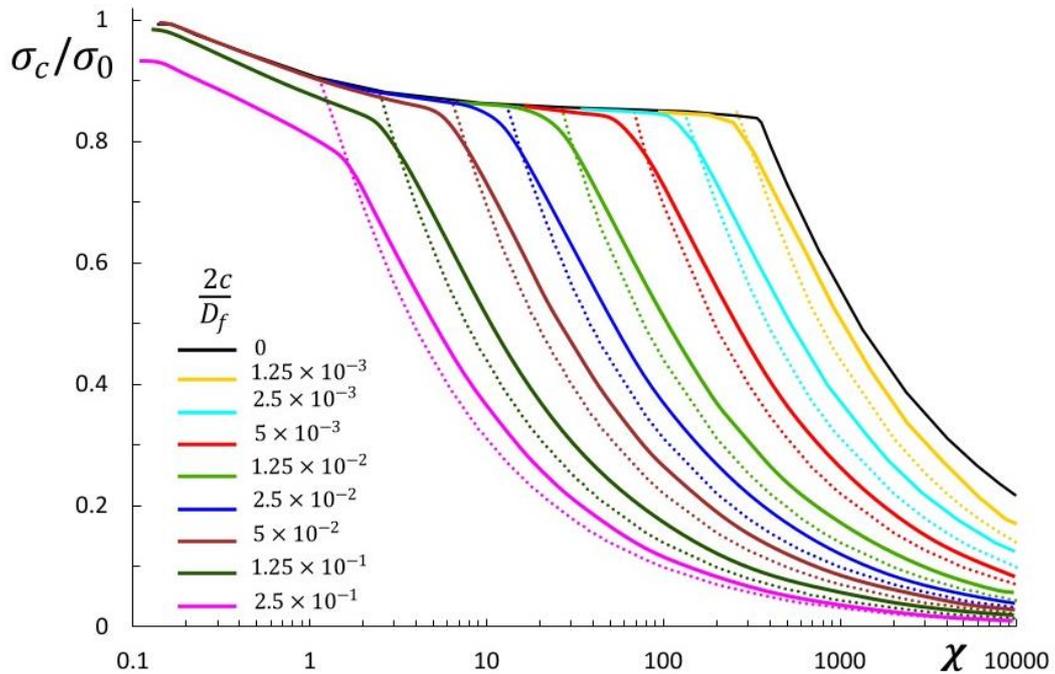

Figure 8. Normalized pull-off strength as a function of $\chi$ for a mushroom fibril with $\beta = 1.2$ and $\xi = 0.01$ with central adhesion defects having various diameters, $2c$. The transition from defect-insensitive to defect-sensitive regimes is observed with increasing $\chi$ and defect size. These results aid in selecting defect-tolerant fibril dimensions for practical applications.

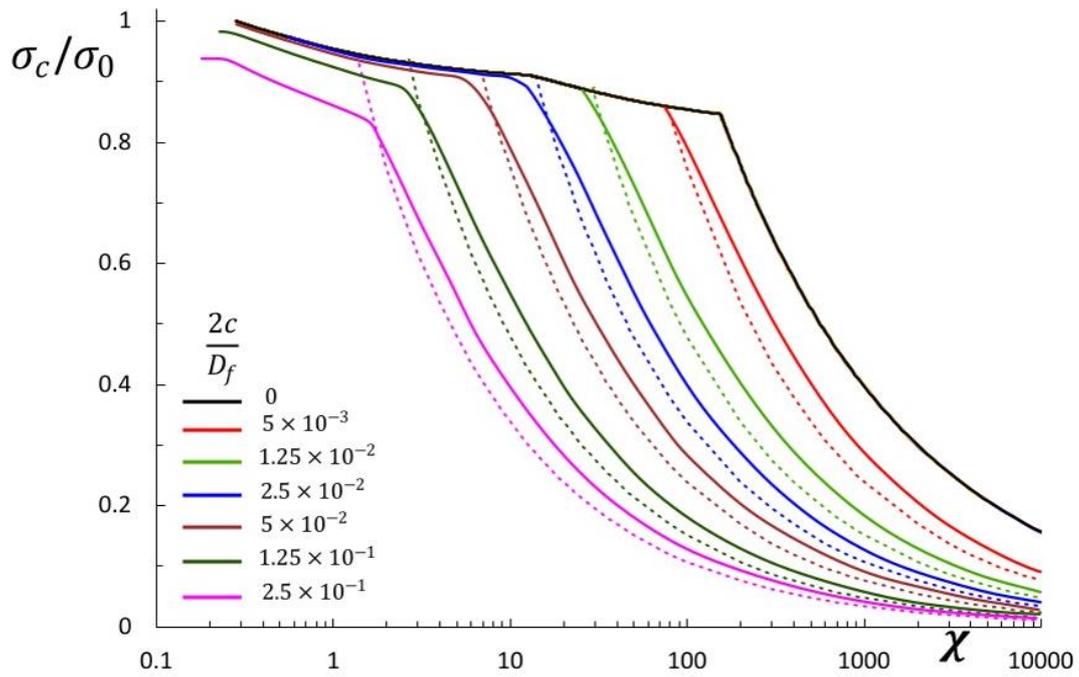

Figure 9. Normalized pull-off strength as a function of $\chi$ for a mushroom fibril with $\beta = 1.15$ and $\xi = 0.01$ with central adhesion defects having various diameters, $2c$. The transition from defect-insensitive to defect-sensitive regimes is observed with increasing $\chi$ and defect size. These results aid in selecting defect-tolerant fibril dimensions for practical applications.

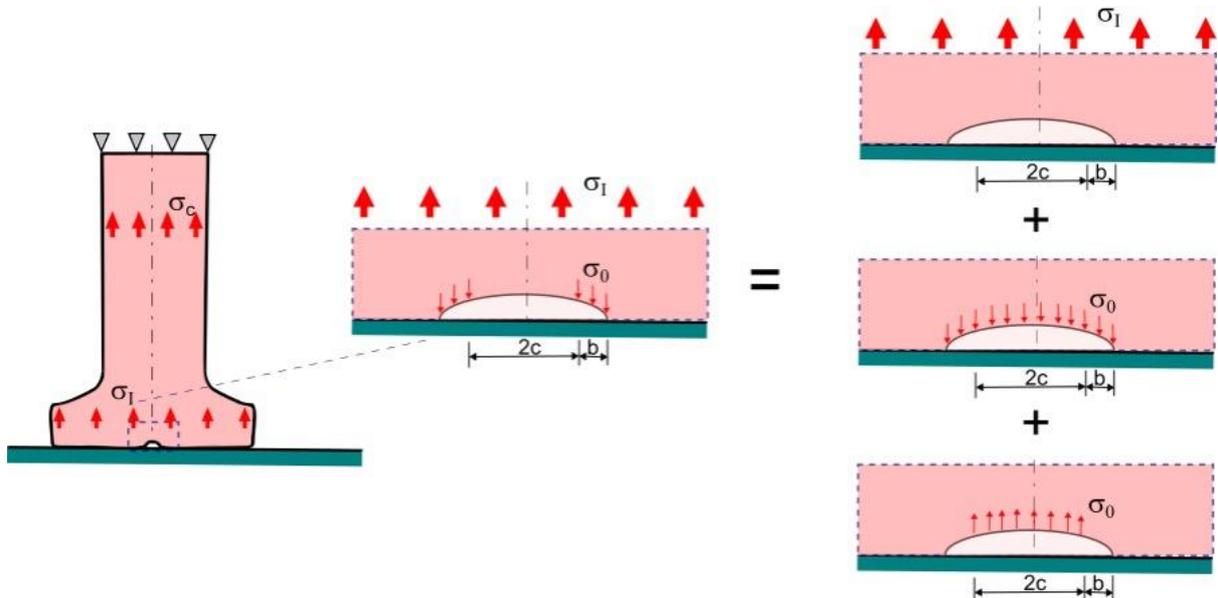

Figure A1. Dugdale cohesive zone for a penny-shaped crack at the substrate-fibril interface.

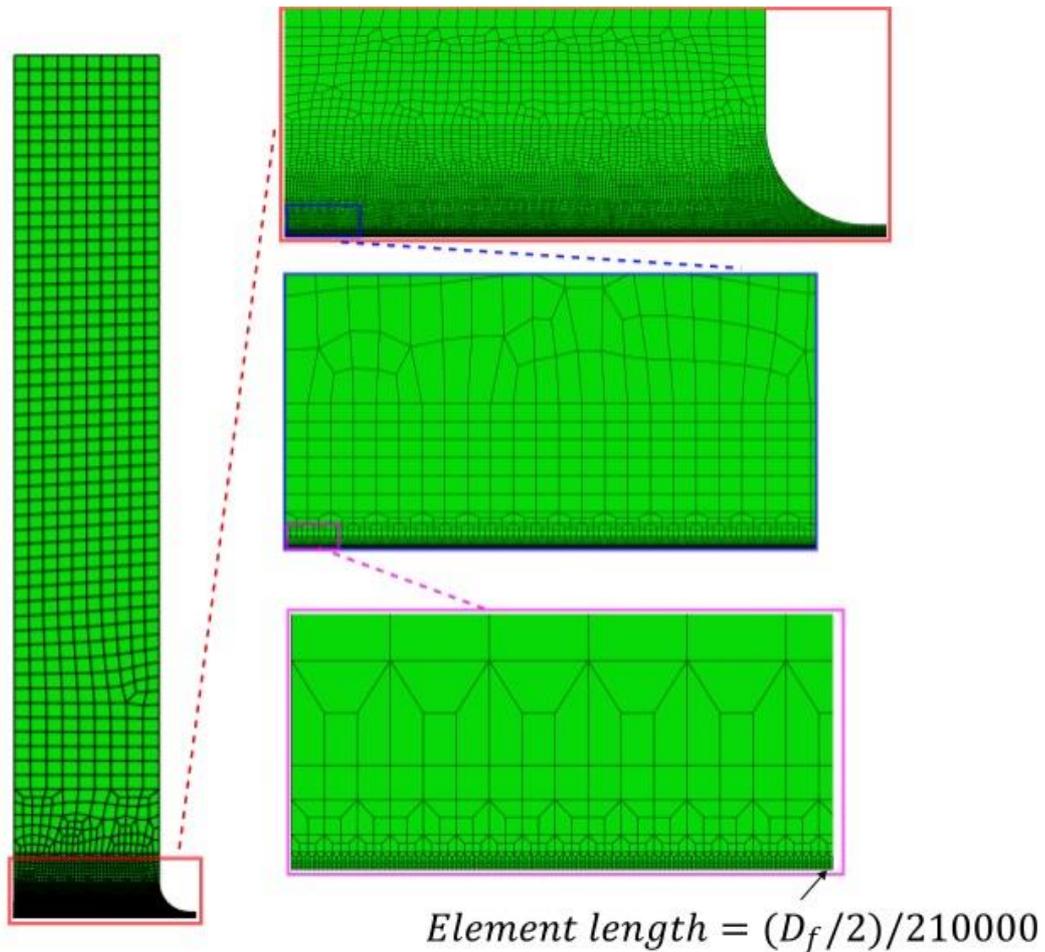

$$Element\ length = (D_f/2)/210000$$

Figure A2. Illustration of a typical finite element mesh.

**Figures captions**

Figure 1. Deformation process for an initially stress free fibril. (a) Geometry of the fibrils; (b) initial deformation state; (c) gradual compression of the mushroom end and development of the Dugdale zone at the center of the fibril-substrate interface and (d) at the perimeter of the fibril mushroom tip.

Figure 2. Load-extension behavior for two mushroom fibril geometries under load control: (a) $\beta = 1.25$ and $\xi = 0.01$, where detachment initiated at the center; (b) $\beta = 1.05$ and $\xi = 0.05$, where detachment initiated at the edge. Each colored triangle denotes the pull-off point for a given value of $\chi$. These curves illustrate the unstable nature of the detachment process and help identify the geometries capable of achieving high pull-off strength through central detachment.

Figure 3. Normalized pull-off strength versus $\chi$ for different mushroom fibril geometries. Each curve corresponds to a ratio of flange to stalk diameter ($\beta$), and each symbol type corresponds to a specific flange thickness ($\xi$). The dashed line represents the performance of a punch geometry for comparison[26]. This figure provides a design map to optimize fibril geometry for desired adhesion performance, highlighting the regimes of flaw insensitivity, gradual degradation, and sharp strength loss.

Figure 4. Normal traction and interface separation along the fibril–substrate interface at the moment of pull-off, for a mushroom fibril with $\beta = 1.25$ and $\xi = 0.01$. Results are shown for multiple values of the parameter $\chi$. Solid lines represent the normal traction; dashed lines represent the separation.

Figure 5. Normal traction and interface separation along the fibril–substrate interface at the moment of pull-off, for a mushroom fibril with $\beta = 1.05$ and $\xi = 0.05$. Results are shown for multiple values of the parameter $\chi$. Solid lines represent the normal traction; dashed lines represent the separation.

Figure 6. Normal traction and interface separation along the fibril–substrate interface at the moment of pull-off, for a mushroom fibril with $\beta = 1.25$ and $\xi = 0.025$. Results are shown for multiple values of the parameter $\chi$. Solid lines represent the normal traction; dashed lines represent the separation.

Figure 7. Effect of a central adhesion defect on the pull-off strength of a mushroom fibril with $\beta = 1.25$ and $\xi = 0.01$. Each curve represents a different defect size, $2c$. Dashed lines show linear elastic fracture mechanics (LEFM) predictions for comparison. This figure quantifies the sensitivity of high-performance fibrils to central defects and underscores the need to consider defect tolerance in adhesive surface design.

Figure 8. Normalized pull-off strength as a function of $\chi$ for a mushroom fibril with $\beta = 1.2$ and $\xi = 0.01$ with central adhesion defects having various diameters, $2c$. The transition from defect-insensitive to defect-sensitive regimes is observed with increasing $\chi$ and defect size. These results aid in selecting defect-tolerant fibril dimensions for practical applications.

Figure 9. Normalized pull-off strength as a function of $\chi$ for a mushroom fibril with $\beta = 1.15$ and $\xi = 0.01$ with central adhesion defects having various diameters, $2c$. The transition from defect-insensitive to defect-sensitive regimes is observed with increasing $\chi$ and defect size. These results aid in selecting defect-tolerant fibril dimensions for practical applications.

Figure A1. Dugdale cohesive zone for a penny-shaped crack at the substrate-fibril interface.

Figure A2. Illustration of a typical finite element mesh.